\documentclass{ws-mpla_arXiv11042022} %%frk: page head removed
\usepackage{graphicx}           %%frk removed mathrsfs,amsmath

\newcommand{\version}{v6}  %%true version v591
\newcommand{\articletype}{Letter}
\newcommand{\beq}{\begin{equation}}
\newcommand{\eeq}{\end{equation}}
\newcommand{\beqa}{\begin{eqnarray}}
\newcommand{\eeqa}{\end{eqnarray}}
\newcommand{\bsubeqs}{\begin{subequations}} %%{\numparts}
\newcommand{\esubeqs}{\end{subequations}} %%{\endnumparts}
\begin{document}

\title{Entropic gravity, minimum temperature, and\\
       modified Newtonian dynamics}

\author{F.R. Klinkhamer}
\address{Institute for Theoretical Physics, University of Karlsruhe,\\
             Karlsruhe Institute of Technology, 76128 Karlsruhe,
             Germany\\
             frans.klinkhamer@kit.edu}

\author{M. Kopp}
\address{Excellence Cluster Universe,
Boltzmannstrasse 2,  85748 Garching, Germany\\
and\\
University Observatory, Ludwig-Maximillians University Munich,  \\
Scheinerstrasse 1, 81679 Munich, Germany\\
michael.kopp@physik.lmu.de}

\maketitle
\begin{abstract}
Verlinde's heuristic argument for the interpretation of the standard
Newtonian gravitational force as an entropic force is generalized by the
introduction of a minimum temperature (or maximum wave length) for the
microscopic degrees of freedom on the holographic screen. With the simplest
possible setup, the resulting gravitational acceleration felt by a test mass $m$
from a point mass $M$ at a distance $R$ is found to be of the form of the
modified Newtonian dynamics (MOND) as suggested by Milgrom. The corresponding
MOND-type acceleration constant is proportional to the minimum temperature,
which can be interpreted as the Unruh temperature of an emerging de-Sitter
space. This provides a possible explanation of the connection between
local MOND-type two-body systems and cosmology.
\vspace*{1\baselineskip}\newline
Journal-ref: \emph{Mod. Phys. Lett. A} \textbf{26}, 2783--2791 (2011)
\vspace*{.5\baselineskip}\newline
Preprint: arXiv:1104.2022\;(\version)
\vspace*{.5\baselineskip}\newline
Keywords: Other theories of gravity, thermodynamics, stellar dynamics and
kinematics.\vspace*{.5\baselineskip}\newline
PACS: 04.50.-h, 05.70.-a, 98.10.+z
\end{abstract}

\maketitle

\newpage
\section{Introduction}
\label{sec:Introduction}

In this \articletype,
we start from Verlinde's heuristic argument\cite{Verlinde2010}
for the standard Newtonian acceleration on a test mass $m$
from an effective point mass $M$ at an effective distance $R$,
the norm of the acceleration three-vector being given by $GM/R^{2}$.
In his approach, classical gravity arises as an
entropic force, hence the name ``entropic gravity.''
Here, we will use
a particular formulation\cite{Klinkhamer2010,Klinkhamer2011}
of Verlinde's argument, which relies only on the Unruh temperature
and holography.

The new ingredient is the introduction of
a minimum temperature $T_\text{min}>0$
for the fundamental microscopic degrees of freedom
on the two-dimensional holographic screen.
The goal of this \articletype~is to explore the consequences
of having this minimum temperature.
Interestingly, we will find that the simplest possible
functional behavior is precisely of the type of Milgrom's
modified Newtonian dynamics (MOND) applied to nonrelativistic
classical gravity.\cite{Milgrom1983,Milgrom1998,McGaugh2011}

It should be mentioned, right from the start, that the key equations of
this article have appeared, in more or less the same form, in the previous
literature. This article is primarily about concepts and logic. For this
reason, the fundamental physical constants, $\hbar$, $c$, and $k_{B}$ are
occasionally displayed, even though typically we use units with
$\hbar=c=k_{B}=1$.

%%\newpage%%tmp
\section{Setup}
\label{sec:Setup}

The $N$ microscopic degrees of freedom on the spherical screen
$\Sigma_{N,\,T,\,T_\text{min}}$
are assumed to be in thermal equilibrium with a temperature%
\bsubeqs\label{eq:T-Tmin-DeltaT}
\beqa
T &=& T_\text{min} +\Delta T\,,
\label{eq:T}\\[2mm]
T_\text{min} &>& 0\,,\quad \Delta T \geq 0.
\eeqa
\esubeqs
An alternative description uses a maximum wavelength $\lambda_\text{max}$
for the thermal excitations (quasiparticles)
of the microscopic degrees of freedom on the holographic screen.
This (reduced) wavelength can be defined as follows:
\beq\label{eq:lambda-max}
c/\lambda_\text{max}
\equiv \frac{1}{2}\; k_B T_\text{min}/\hbar\,.
\eeq
Furthermore, the setup requires the following behavior
for the macroscopic variables corresponding to the effective mass $M$
and the area $A_\Sigma\equiv 4\pi R^{2}$:
\beq\label{eq:DeltaT-M-conditions}
M        \propto  N\,\Delta T \,,\quad
A_\Sigma \propto N \,,
\eeq
as will be discussed in Sec.~\ref{sec:Heuristic-argument}.

The physical picture, now, is as follows.
Having $T_\text{min} > 0$ for the microscopic degrees of freedom
of a given (inner) holographic screen $\Sigma_{N,\,T,\,T_\text{min}}$
corresponds to having a nonzero entropy $S_\text{min} > 0$.
Such a nonzero entropy can be \emph{interpreted}
as being due to missing information\cite{Hawking1996} from
the presence of an event horizon for the degrees of freedom
on the inner screen (``observers'' in the usual terminology).
 From the holographic point of view,\cite{Verlinde2010}
having a  maximum wavelength $\lambda_\text{max}$ for the microscopic
degrees of freedom on the screen is certainly consistent
with obtaining a finite
length scale in the emerged space.\footnote{The event horizon
can perhaps also be interpreted as an (outer) holographic
screen $\Sigma_\text{out}$.
It appears that the correct description is then that \emph{each}
holographic screen, $\Sigma_{N,\,T,\,T_\text{min}}$ or $\Sigma_\text{out}$,
has its \emph{own} emerged space
(a similar point has been made by Penrose\cite{Penrose1996}
in an entirely different context).
Still, in order to describe the behavior of the test mass $m$
near the inner screen, it may turn out to be useful to work
in some type of ``average space'' between the two
surfaces.\label{ftn:compromise-space}}

\begin{figure}[t]   %%twocolumn
%
%\begin{figure*}[p]   %%preprint
%\begin{center}
\hspace*{-0.4cm}
\includegraphics[width=13.4cm]
{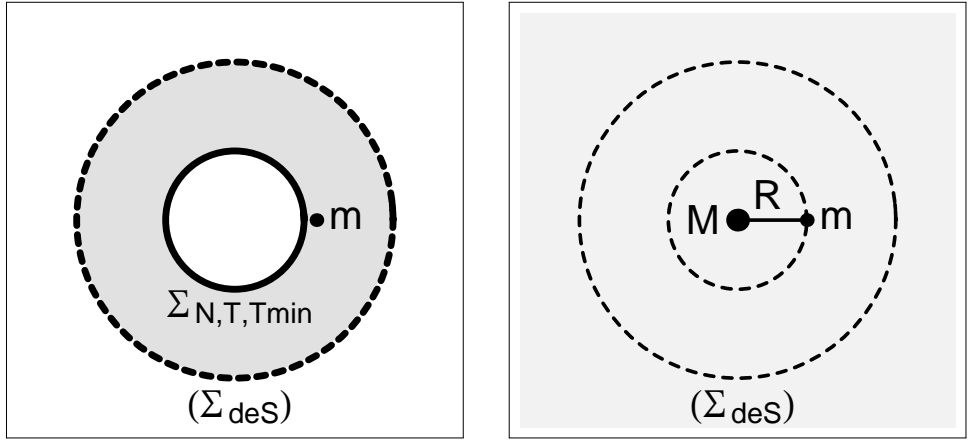}%% fig1_v424bis --> fig1_v5
%{entropic-grav-mond_fig1_v424bis.eps}
%\end{center}
\vspace*{-0cm}
\caption{Left panel: test mass $m$ at rest
in the emerged space (dark shading), just outside
the spherical holographic screen $\Sigma_{N,\,T,\,T_\text{min}}$
(full heavy curve).
The fundamental microscopic degrees of freedom of the
screen $\Sigma_{N,\,T,\,T_\text{min}}$ have a minimum temperature $T_\text{min}$
and a corresponding event horizon (dashed heavy curve), which can
possibly be identified as the de-Sitter horizon $\Sigma_\text{deS}$.
Right panel: gravitational attraction experienced by the
test mass $m$, as coming from
an effective point mass $M\propto N\,(T-T_\text{min})$
at an effective distance $R\propto N^{1/2}$
in a standard spacetime (light shading), possibly de-Sitter spacetime.
\vspace*{-0cm}}
\label{fig:emerged-space}
\end{figure}

For the physics near the inner screen, it is important to
understand that the event horizon is a derived effect and
that what really matters is the heat-bath-type temperature
$T_\text{min}$ of the microscopic degrees of freedom on the
holographic screen; see the left panel of Fig.~\ref{fig:emerged-space}.
The extra energy from an additional temperature $\Delta T$
of the degrees of freedom on the inner screen
is responsible for a net attraction on a stationary
test mass $m$ just outside the screen
(see Sec.~\ref{sec:Heuristic-argument} for details).
According to Verlinde,\cite{Verlinde2010}
the resulting gravitational force $\mathbf{F}_\text{grav}$
on a test mass $m$ can be \emph{interpreted} as coming
from an effective point mass $M$ at an effective distance $R$
in an effective geodesically-complete spacetime;
see the right panel of Fig.~\ref{fig:emerged-space}.
It needs to be emphasized that the right panel
of Fig.~\ref{fig:emerged-space} is now considered to give only an
approximate and derived description of the ``physical reality,''
whereas the left panel is taken to give a more accurate
and more fundamental description.

%%\newpage%%tmp
\section{De-Sitter realization}
\label{sec:DeS-realization}

In the previous section, we have argued that
the existence of an intrinsic minimum temperature $T_\text{min}$
for the degrees of freedom of the inner screen
corresponds to the presence of an effective event horizon
for these degrees of freedom. Now, identify this effective event
horizon with the event horizon $\Sigma_\text{deS}$
in an emerged de-Sitter (deS) space, so that $T_\text{min}$
equals the corresponding Unruh
temperature.\cite{Unruh1976,GibbonsHawking1977,BellLeinaas1982,Narnhofer-etal1996,Acquaviva-etal2011}
With the Gibbons--Hawking result\cite{GibbonsHawking1977}
$T_\text{deS} = H_\text{deS}/(2\pi)$ for a spherical event horizon
at $r=c/H_\text{deS}$ in a static de-Sitter metric, we then have
\beq\label{eq:Tmin-deS}
2\pi\,T_\text{min} = H_\text{deS} \equiv H\,,
\eeq
where $H$ is a useful short-hand notation.\footnote{An early
paper\cite{Easson-etal2010} on entropic gravity in a cosmological context also
discusses a minimum temperature, but the setup of that paper
is different from the one presented here.\label{ftn:early-paper}}

Next, examine a detector with uniform linear
acceleration $\mathbf{A}$ in de-Sitter space. The resulting Unruh-type
temperature has been calculated in Ref.~\refcite{Narnhofer-etal1996}:
$(2\pi \,T)^{2}=|\mathbf{A}|^{2}+H^{2}$.
Inverting this result, in the spirit of Ref.~\refcite{Verlinde2010},
and using \eqref{eq:T} and \eqref{eq:Tmin-deS} gives
\beq\label{eq:A-deS}
|\mathbf{A}| = \sqrt{(2\pi \,T)^{2} -H^{2}} =
       2\pi\,\Delta T\,\sqrt{1+2\,T_\text{min}/\Delta T}\,.
\eeq
The first equality in \eqref{eq:A-deS} can be understood as the
correction to the acceleration associated
with a local temperature $T$ if Minkowski spacetime is replaced
by de-Sitter spacetime (which has an event horizon
even if the acceleration of the detector vanishes).

 From \eqref{eq:A-deS}, a quadratic equation in $\Delta T$ is obtained,
which has the following positive root:
\beqa\label{eq:DeltaT-deS}
2\pi\,\Delta T
&=&
\sqrt{|\mathbf{A}|^{2}+(2\pi\,T_\text{min})^{2}}-2\pi\,T_\text{min}\,.
\eeqa
For our purpose, de-Sitter space is only an auxiliary ingredient
and we continue to work with the expression \eqref{eq:DeltaT-deS}, solely 2
defined in terms of
$T_\text{min}$ from the holographic
screen.\footnote{Observe that the right-hand side of \eqref{eq:DeltaT-deS}
is the simplest
possible function of $|\mathbf{A}|$ which reduces to $|\mathbf{A}|$ for
$T_\text{min}=0$, drops to $0$ for $T_\text{min}\to\infty$,
and involves $|\mathbf{A}|$ only in
the combination $|\mathbf{A}|^{2}+(2\pi\,T_\text{min})^{2}$.
}

Still, de-Sitter space is special, because the Unruh-type
temperature $T$ for a uniform linear acceleration of
the detector is invariant under local Lorentz transformations
of the detector motion.\cite{Narnhofer-etal1996}
The surprising role of special relativity in the
Verlinde-type `derivation' of standard Newtonian gravity
has already been noted in Ref.~\refcite{Klinkhamer2010}.
Apparently, the importance of local Lorentz invariance
also holds for the `derivation' of modified Newtonian gravity
(see Sec.~\ref{sec:Heuristic-argument}).

It may be that, for the case of a holographic screen with
minimum temperature $T_\text{min}$, the demand of local Lorentz
invariance uniquely selects a de-Sitter space
with a Hubble constant given by \eqref{eq:Tmin-deS}.
But, for now, we simply \emph{assume} de-Sitter space to be relevant or,
at least, to provide a good approximation for the physics investigated.

%%\newpage%%tmp
\section{Heuristic argument}
\label{sec:Heuristic-argument}

At last, we are ready to calculate the gravitational
attraction experienced by a stationary test mass $m$ just outside the holographic screen
$\Sigma_{N,\,T,\,T_\text{min}}$ as shown in the left panel of Fig.~\ref{fig:emerged-space}.
The procedure is simple: reverse \eqref{eq:DeltaT-deS}
and use, starting from $2\pi\,\Delta T$, the Verlinde-type argument
as given in Eq.~(4) of Ref.~\refcite{Klinkhamer2010}. The norm of the
inward radial acceleration $\mathbf{A}$ of the test mass $m$
generated by the screen quantities $N$, $T$, and $T_\text{min}$
is then found to be given by the following expression:
\bsubeqs\label{eq:main-results}
\beq\label{eq:main-result-MONDeq}
|\mathbf{A}|\;\widehat{\mu}
\left(\frac{|\mathbf{A}|}{4\pi c\; k_B T_\text{min}/\hbar}\right)
=2\pi c\;k_{B}\Delta T/\hbar
= GM/R^{2}\,,
\eeq
with fundamental constants $\hbar$, $c$, and $k_{B}$ restored
and with definitions
\beqa
\widehat{\mu}(x)&\equiv& \sqrt{1+1/(2x)^{2}} -1/(2x)  \,,
\label{eq:main-result-widehatmu}\\[2mm]
G    &\equiv&  f\,c^3\,l^{2}/\hbar\,,
\label{eq:main-result-G}\\[2mm]
M  &\equiv&  \frac{1}{2}\,N\,k_{B}\Delta T/c^{2}\,,
\label{eq:main-result-M}\\[2mm]
A_\Sigma    &\equiv&  f\,N\,l^{2}\,,
\label{eq:main-result-ASigma}\\[2mm]
R^{2}  &\equiv&
\frac{1}{4\pi}\; A_\Sigma\,.
\label{eq:main-result-R2}
\eeqa
\esubeqs

Strictly speaking, the last step of  `derivation'
\eqref{eq:main-result-MONDeq} is trivial, as it involves
only mathematical definitions, \textit{viz.}
Eqs.~\eqref{eq:main-result-G}--\eqref{eq:main-result-R2}.
The real issue is, of course, to establish the corresponding
physical picture. We start with six technical comments
and, then, follow-up with a few general remarks.
In a first reading, it is possible to skip
these clarifications and to proceed directly to
Sec.~\ref{sec:Discussion}.

%%\newpage%%tmp
First, the quantity $l^{2}$ entering \eqref{eq:main-result-G}
may (or may not) correspond to a new fundamental constant of nature,
the quantum of area.\cite{Klinkhamer2010,Klinkhamer2011}
The quantity $f$ in \eqref{eq:main-result-G} is then an appropriate
numerical factor appearing from the calculation of $G$
(for $f=1$, the length $l$ equals the standard Planck length scale).

Second, macroscopic quantities
in \eqref{eq:main-result-G}--\eqref{eq:main-result-R2} are denoted by
upper-case letters and fundamental constants by lower-case letters.
More specifically, $G$, $M$, and $R^{2}$ are \emph{effective}
macroscopic quantities, derived from the fundamental quantities
$N$ and $\Delta T$ describing the microscopic
degrees of freedom on the holographic screen.

Third, the behavior $N\propto A_\Sigma$ from \eqref{eq:main-result-ASigma}
corresponds to holography; see Ref.~\refcite{Verlinde2010} for further
discussion and references. The crucial assumption, here,
is that $N$ is a purely geometric quantity, that is, $N$ is dependent
on the area but not on the temperature
($N$ is, for example, not proportional to the combination
$A_\Sigma\,\Delta T/T$).

Fourth, given the number $N$ of degrees of freedom on the screen,
the extra energy $\textstyle{\frac{1}{2}}\,k_{B}\Delta T$ per degree
of freedom provides for an acceleration of the test mass $m$,
which is absent in the perfect (empty, matter-free)
de-Sitter space with $T = T_\text{min}$ on the screen.
In this way, it makes sense that the effective Newtonian
mass $M$ is defined to be proportional to $N$ and $\Delta T$,
as shown by \eqref{eq:main-result-M}.
In fact, it is possible to imagine that
the holographic screen consists of a gas of
nonrelativistic ``atoms of two-dimensional space.''
The velocities of these identical atoms,
$\{ \mathbf{u}_{n}=\mathbf{v}_{n}
                   +\mathbf{w}_{n}\,|\, n=1,\,\dots\,,\, N/2\}$,
are assumed to be built from two sets of independent random
velocities, $\{\mathbf{v}_{n} \}$ and $\{  \mathbf{w}_{n} \}$,
which give rise to $T_\text{min}$ and $\Delta T$, respectively.
The kinetic energy of the second set of random velocities,
$\{ \mathbf{w}_{n} \}$,
then corresponds to the effective Newtonian mass $M$. Note that the
corresponding gravitational force \eqref{eq:main-result-MONDeq}
is not quite a standard entropic force (having $|\mathbf{F}|\propto T$)
but a modified entropic force with a shifted temperature scale
(having $|\mathbf{F}|\propto \Delta T\equiv T-T_\text{min}$).

Fifth, it is possible to generalize the argument used in
Eqs.~\eqref{eq:main-result-MONDeq}--\eqref{eq:main-result-R2}
by allowing for modifications
of the energy equipartition law of the microscopic
degrees of freedom,\cite{Klinkhamer2011}
but this is not necessary for the present discussion.

Sixth, an alternative `derivation' of \eqref{eq:main-result-MONDeq}
which directly starts from Verlinde's entropic-force formula
is given in the Appendix.

%%\newpage%%tmp
We now present the promised general remarks, intended to
further clarify the physical picture
(see Ref.~\refcite{Verlinde2010} for additional details).
These remarks are primarily concerned with the emergent space
from the holographic screen and are highly speculative,
because the fundamental theory is unknown
(the ultimate goal is, of course, to learn something
about this fundamental theory, a first clue perhaps having been
found in Ref.~\refcite{Klinkhamer2011}).

By increasing or reducing the number $N$ of degrees of freedom
on the holographic screen the effective distance $R$ between
the masses $M$ and $m$ grows or shrinks, according to
\eqref{eq:main-result-ASigma}--\eqref{eq:main-result-R2}. In fact, reducing
$N$ corresponds to a coarse-graining of the degrees of freedom
(similar to Kadanoff's block-spin transformation in lattice models)
and the resulting information
(new coupling constants in the effective theory
coming from the block-spinning) corresponds to an increased range of the
orthogonal space coordinates, consistent with the picture
of a shrinking surface at the inner boundary of the emerged space
in the left panel of Fig.~\ref{fig:emerged-space}.
For the present setup, the maximally coarse-grained surface
is the Schwarzschild horizon.\cite{Verlinde2010}

Increasing $N$, while keeping $M$ fixed,
moves the screen out towards the de-Sitter horizon
and the screen temperature $T$ approaches $T_\text{min}$ from above,
according to \eqref{eq:main-result-M}. However,
as discussed in Footnote~\ref{ftn:compromise-space}, the naive
description in terms of a single emerged space can be expected
to become invalid as the inner screen approaches the outer one.

%%\newpage%%tmp
\section{Discussion}
\label{sec:Discussion}

The first equality in \eqref{eq:main-result-MONDeq} already appears
in a prescient  paper by Milgrom,\cite{Milgrom1998}
but the heuristic `derivation' of the second equality is new
and really makes for
MOND applied to nonrelativistic classical gravity.\cite{Milgrom1983}
The crucial extra input compared to Ref.~\refcite{Milgrom1998} is
the combination \eqref{eq:main-result-M} and  \eqref{eq:main-result-ASigma},
see also the third and fourth technical comments in the previous section.

 From the heuristic argument of the previous section or the one of the
Appendix, the gravitational
attraction of a stationary test mass $m$ to a point mass $M$
at a distance $R$ (right panel of Fig.~\ref{fig:emerged-space})
is thus found to give the following inward acceleration $\mathbf{A}$
of the test mass $m$:
\bsubeqs\label{eq:MONDeq-A0}
\beq\label{eq:MONDeq}
\mathbf{A}\;\,\widehat{\mu}\,\big(|\mathbf{A}|/A_{0}\big)=
-\big(GM/R^{2}\big)\,
\widehat{\boldsymbol{n}}\,,
\eeq
with $\widehat{\boldsymbol{n}}$ a unit vector
pointing from $M$ to $m$, the explicit function $\widehat{\mu}(x)$
from \eqref{eq:main-result-widehatmu},
having $\widehat{\mu}(x) \to 1$ for $x\to\infty$  and
$\widehat{\mu}(x) \to x$ for $x\to 0$, and the acceleration constant
\beq\label{eq:A0}
A_{0} = 4\pi c\; k_B T_\text{min}/\hbar
= 8\pi c^{2}/\lambda_\text{max}\,,
\eeq
in terms of the maximum wavelength defined by \eqref{eq:lambda-max}.
As explained in Sec.~\ref{sec:DeS-realization},
an effective de-Sitter space
has been assumed to be relevant for the type of holographic screen
considered and the corresponding horizon distance is given by
\beq
c/H_\text{deS} = \hbar\, c/(2\pi\, k_B T_\text{min})
= 1/(4\pi)\,\lambda_\text{max}\,.
\eeq
\esubeqs
Eliminating $T_\text{min}$ (or $\lambda_\text{max}$) from
the last two equations gives
\beq\label{eq:A0-HdeS}
A_{0} = 2\, c\;H_\text{deS}\,,
\eeq
which will be discussed later.

Note that \eqref{eq:MONDeq} can be expected to hold
for linear motion ($m$ moving towards or away from $M$) but
not for circular motion ($m$ orbiting $M$), relevant to the
rotation curves of galaxies.\cite{Milgrom1983} The constant
$a_{0} \approx 1.2\times 10^{-8}\;\text{cm}\:\text{s}^{-2}$
obtained from the best available rotation-curve data\cite{McGaugh2011}
can be expected to differ from
our $A_{0}$ by a factor of order unity.\cite{BellLeinaas1982}
In addition, \eqref{eq:A0-HdeS} is considered to hold for
an exact de-Sitter space, but the present
universe is not a perfect de-Sitter space, which
will slightly change the temperature
formula \eqref{eq:A-deS} and, thus, the resulting value of
$A_{0}$.\cite{Acquaviva-etal2011}
Still, the order of magnitude of $A_0$ from \eqref{eq:A0-HdeS}
is quite reasonable,
$A_{0} \sim 10^{-7}\;\text{cm}\;\text{s}^{-2}$,
if $H_\text{deS}$ is identified with
$\sqrt{3/4}\approx 0.87$ times the measured Hubble constant
$H_{0}\approx 75\;\text{km}\:\text{s}^{-1}\:\text{Mpc}^{-1}$
[the square root factor follows from the standard Friedmann
equation of a spatially flat Universe with
energy density ratio $\rho_\text{vacuum}/\rho_\text{matter} = 3$].

The possible relation of entropic gravity and
MOND has been discussed in several recent papers; see, e.g.,
Refs.~\refcite{LiChang2010,HoMinicNg2010,KiselevTimofeev2010,Neto2010,Pikhitsa2010}.
Directly relevant to our discussion is the paper
by Pikhitsa,\cite{Pikhitsa2010} of which
we only became aware when writing up this \articletype.
Not surprisingly, his basic equations are the
same as ours, but the precise claims
and physical interpretation are different. For example,
we do not claim to have obtained the MOND acceleration constant
$a_{0}$ relevant for circular
motion. And our direct physical interpretation of
\eqref{eq:main-result-M} does not rely upon results
from general relativity as appears to be the case for
Eqs.~(4)--(5) in Ref.~\refcite{Pikhitsa2010}.
Still, our main physical conclusion is the same as Pikhitsa's,
namely, that MOND may be related
to the existence of a minimum temperature. However, in
the spirit of Verlinde's approach,\cite{Verlinde2010}
we reverse cause and effect:
a minimum temperature (maximum wave length) of the fundamental
microscopic degrees of freedom responsible for
classical gravity may produce a MOND-like
behavior at sufficiently small accelerations of the test mass.

The question arises as to the nature of the holographic
screen if the minimum temperature of its degrees of freedom
is indeed nonzero. One possible explanation is that
these fundamental microscopic degrees of freedom
of the screen are in a long-lived metastable state.
(Having such a metastable state may not be altogether unreasonable
if the microscopic degrees of freedom have long-range interactions as
has been argued to be the case in Ref.~\refcite{Klinkhamer2011}.)
If the interpretation as a metastable state is correct, then
there is, in principle, the possibility of a discontinuous reduction of
the MOND-type acceleration constant \eqref{eq:A0}.
In turn, this may lead to a discontinuous decay of the
corresponding de-Sitter spacetime (it is not clear if
there is any relation with the type of de-Sitter decay
recently discussed by Polyakov\cite{Polyakov2009}).

Let us, finally, return to \eqref{eq:A0-HdeS}, which relates
a characteristic, $A_{0}$, of small-scale two-body
dynamics \eqref{eq:MONDeq} to a cosmological quantity, $H_\text{deS}$.
\textit{A priori,} such a relation would be hard to understand.
But this article suggests that both quantities
(the ``local'' $A_{0}$  and the ``global'' $H_\text{deS}$)
have a \emph{common origin}.
As shown by the left panel of Fig.~\ref{fig:emerged-space},
the suggestion is that a minimum temperature $T_\text{min}$
(or maximum wave length $\lambda_\text{max}$) of the fundamental
microscopic degrees of freedom on the holographic screen gives rise to
both the MOND-type acceleration constant $A_{0} \sim T_\text{min}$
and the de-Sitter horizon distance $1/H_\text{deS}\sim 1/T_\text{min}$.

Granting the approximate equality both of $A_{0}$ and the inferred
MOND acceleration constant $a_{0}$ for circular motion
and of $H_\text{deS}$ and the measured Hubble constant $H_{0}$ from
the expanding Universe, there is then a logical connection between
$a_{0}$ and  $H_{0}$, resulting in the relation $a_{0} \sim c\,H_{0}$
(the approximate numerical coincidence of $a_{0}$ and $c\,H_{0}$ was
already noted in Milgrom's original paper\cite{Milgrom1983}).
The logical connection between $a_{0}$ and $c\,H_{0}$
is indirect, as each of them traces back to the apparently
more fundamental quantity $T_\text{min}$.
Of course, all this only makes sense if the heuristic argument used
here is physically relevant.

%%\newpage%%tmp
\begin{appendix}
\section{Alternative heuristic argument}
\label{sec:Appendix}
%%
%% Authors' Note: We want eqs. in appendix numbered as (A.1), (A.2), etc
%%                Seems to be automatic!
%%

In this appendix, the main result in \eqref{eq:main-result-MONDeq}
is obtained by directly following Verlinde's original
argument.\cite{Verlinde2010} The starting point is
the entropic-force formula ($\hbar=c=k_{B}=1$)
\beq\label{eq:Fentropic}
\mathbf{F}_\text{grav}=\Big[\,T\,\boldsymbol{\nabla}S\,\Big]_{\Sigma_{0}}\,,
\eeq
for a single spherical holographic screen $\Sigma_{0}$ with area
$4\pi R^2 \propto N$
and effective mass $M_{0}=\textstyle{\frac{1}{2}}\,N\,T$,
combined with the assumption
\beq\label{eq:S-assumption}
\boldsymbol{\nabla}S =-2\pi\, m\; \widehat{\boldsymbol{n}}_{0}\,,
\eeq
for unit normal $\widehat{\boldsymbol{n}}_{0}$ of the screen $\Sigma_{0}$
directed towards the particle with mass $m$
(the particle is separated from the screen by a distance of the order
of its Compton wave length, $\hbar/mc$).

The basic idea, now, is to replace the original entropic-force formula
\eqref{eq:Fentropic} by
\beq\label{eq:Fentropic-general}
\mathbf{F}_\text{grav} =
\sum_{n=1}^{2K} \,
\Big[\,T\,\boldsymbol{\nabla}S \,\Big]_{\Sigma_{n}}\,,
\eeq
where the sum includes three types of contributions and has
an integer $K\gg 1$ to control the number of terms.
The first type of contribution in \eqref{eq:Fentropic-general}
comes from the
main spherical screen $\Sigma_{N,\,T,\,T_\text{min}} \equiv \Sigma_{1}$
discussed in Sec.~\ref{sec:Setup}.
The second type of contribution comes from
a plane screen $\Sigma_{2}$ with $T=T_\text{min}$, where $\Sigma_{2}$ is
orthogonal to the $\widehat{\boldsymbol{n}}_{1}$ normal from $\Sigma_{1}$
passing through the particle (specifically,
$\widehat{\boldsymbol{n}}_{2}=-\widehat{\boldsymbol{n}}_{1}$)
and $\Sigma_{2}$ is positioned on the other side of the particle $m$
compared to $\Sigma_{1}$.
The third type of contribution comes from many ($K-1\gg 1$) pairs
of parallel plane $T=T_\text{min}$ screens having
different random orientations
($\widehat{\boldsymbol{n}}_{n} \times \widehat{\boldsymbol{n}}_{1} \ne 0$
for $n\geq 3$) and sandwiching the particle between them.
For simplicity, the pure de-Sitter screens have been taken
to be infinite planes, rather than spheres with very large radii
($c/H_\text{deS} \gg R$).
The particle is thus surrounded by $2K-1$ plane screens $\Sigma_{n}$
(for $n=2,\,\ldots \,,\,2K$) with temperature $T_\text{min}$
and a single spherical screen $\Sigma_{1}$
with temperature $T\geq T_\text{min}$.
The corresponding physical picture is effectively
that of a particle $m$ immersed in an anisotropic heat bath
due to de-Sitter space and the localized energy density.

Using \eqref{eq:S-assumption} and the de-Sitter-space
Unruh temperatures,\cite{GibbonsHawking1977,Narnhofer-etal1996} all
matching contributions from pure de-Sitter screens
cancel in the sum \eqref{eq:Fentropic-general}
and we are left with only two contributions
(from the screens $\Sigma_{1}$ and $\Sigma_{2}$):
\beq\label{eq:Fentropic-two-screens}
\mathbf{F}_\text{grav}=
- m\,\Big(\sqrt{|\mathbf{A}|^2+(2\pi\,T_\text{min})^2}
          -2\pi\,T_\text{min}\Big)\,
\widehat{\boldsymbol{n}}_{1}\,.
\eeq
Simply taking over Verlinde's `derivation'
(Sec.~3.2 of Ref.~\refcite{Verlinde2010}) of the standard Newtonian
gravitational force $|\mathbf{F}_{\text{grav},\,0}|=m\,GM_{0}/R^2$
for the $T_\text{min}=0$ case (corresponding to Minkowski spacetime)
gives for the norm of \eqref{eq:Fentropic-two-screens} in reversed order:
\beq\label{eq:Fentropic-alternative}
m\,\Big(\sqrt{|\mathbf{A}|^2+(2\pi\,T_\text{min})^2}
        -2\pi\,T_\text{min}\Big)
=
m\,\big(GM_{0}/R^2\big)\;\Big(1+\text{O}\big(T_\text{min}/T\big)\Big)\,,
\eeq
which, to leading order in $T_\text{min}/T$,
reproduces the behavior of \eqref{eq:main-result-MONDeq}.
Note that the mass $M_{0}$ times the last factor  2
in brackets of \eqref{eq:Fentropic-alternative} corresponds to the mass $M$
defined by \eqref{eq:main-result-M}.

\end{appendix}

%%%%\newpage%%tmp
\hfill
\section*{Acknowledgements}
FRK thanks M. Milgrom for valuable discussions over the years.

%%\newpage%%tmp

\end{document}